\documentclass[submission]{eptcs}

\usepackage{color}
\usepackage{graphicx}
\usepackage{booktabs}
\usepackage{subfig}

\providecommand{\event}{FMSPLE 2016} 

\title{Incremental Consistency Checking in Delta-oriented UML-Models for Automation Systems}
\author{Matthias Kowal
\institute{TU Braunschweig\\ Braunschweig, Germany}
\email{m.kowal@tu-braunschweig.de}
\and
Ina Schaefer
\institute{TU Braunschweig\\ Braunschweig, Germany}
\email{{i.schaefer@tu-braunschweig.de}
}}
\def\titlerunning{Consistency Checking in Delta-oriented UML-Models}
\def\authorrunning{M. Kowal \& I. Schaefer}
\begin{document}
\maketitle

\begin{abstract}
Automation systems exist in many variants and may evolve over time in order to deal with different environment contexts or to fulfill changing customer requirements. This induces an increased complexity during design-time as well as tedious maintenance efforts. We already proposed a multi-perspective modeling approach to improve the development of such systems. It operates on different levels of abstraction by using well-known UML-models with activity, composite structure and state chart models. Each perspective was enriched with delta modeling to manage variability and evolution. As an extension, we now focus on the development of an efficient consistency checking method at several levels to ensure valid variants of the automation system. Consistency checking must be provided for each perspective in isolation, in-between the perspectives as well as after the application of a delta.    
\end{abstract}

\section{Introduction}
Automation systems are typically built to operate several years or even decades~\cite{zvei}. The long lifetime poses a challenge in two different ways: First, all hardware components possess a limited durability and identical spare parts may not be available for the full lifetime of the automation system. Thus, alternative spare parts have to be installed, which may affect the hardware and the software of the remaining system, representing an unanticipated change or evolution in the system. Second, we have to manage variability in form of multiple environment contexts such as legal requirements or differing customers~\cite{LG+2012,HW+2013}. For instance, customer X requires a faster conveyor line than costumer Y which means that the hardware and software setup have to be adapted. Both aspects, variability and evolution, lead to a high complexity for the system design, an increased maintenance effort and the impression to view automation systems as a product line~\cite{BB+2012}.             

By considering an automation system as a product line, we already proposed a design-level modeling approach following the separation of concerns principle. We introduced three perspectives based on well-known UML-models. Each perspective provides a different view onto the system. First, the workflow perspective shows the paths of a workpiece throughout the system, e.g., which conveyor belt a piece will take. Second, hardware elements, e.g., robots and sensors, are abstracted as architectural components in the architectural perspective. The behavioral perspective captures the implementation of the architectural components by state-based models. Entities from the individual perspectives can be mapped on each other to fully represent a single variant of an automation system. Tasks in the workflow are connected to components and each component is mapped to a state chart. In the next step, the three perspectives as well as the mapping are enhanced with delta modeling to manage variability and evolution~\cite{InaDelta} and therefore fully support the development of an automation system as a product line. In delta modeling, we have a core model and multiple deltas encapsulating modifications. A new system variant is obtained by applying a set of deltas to the core model resulting in the desired variant. In a last step, we are able to automatically generate source code for the modeled automation system. The approach is validated with a laboratory case example~\cite{icse-ws,KowalAT2014}.            

This work is an extension of the multi-perspective modeling approach~\cite{icse-ws,KowalAT2014} considering the aspect of consistency. In our previous work, consistency checking was a manual and tedious task executed by the developer. Consistency must be ensured on three separate levels leading to a high complexity and making it almost impossible for humans in large-scale automation systems. Each perspective must be consistent in isolation (\textit{Level 1}) as well as consistency must be ensured in-between the different perspectives (\textit{Level 2}). In addition, we must ensure a valid variant after the application of a delta (\textit{Level 3}) which may concern all perspectives as well as the mapping between the perspectives. Especially, \textit{Level 3} is challenging due to an increasing number of consistency checks, which is, of course, directly related to the number of variants in the product line. This problem is further emphasized by large models leading to a massive amount of redundant consistency checks if we have almost identical variants. To tackle this problem, in this paper we propose an efficient consistency checking method based on our previous work~\cite{icse-ws,KowalAT2014}.   

The paper outline is as follows: Section~2 describes the multi-perspective modeling approach in more detail. In Section~3, we introduce three different consistency checking concepts that are applicable for our approach. A description of the case example and its evaluation can be found in Section~4. The prototypical implementation is elaborated in Section~5. Section~6 covers related work, followed by concluding remarks and future work in Section~7.

\section{Multi-perspective Modeling for Automation Systems}
Model-driven development (MDD) has several benefits compared to code-based software system development, which makes it increasingly more interesting for industrial use cases~\cite{mellor2003model,ibm,Vya2013,BS+2011,KP+2007}. Especially considering large scale software systems, one can observe a rising productivity with MDD after a first adjustment phase. In particular, the expected advantages are as follows~\cite{primerMDA,atkins,rumpe,schmidt}:   

\begin{itemize}
\item Models provide a higher level of abstraction resulting in an easier and better understandable way to describe the system during all development phases, since we can hide complex details.
\item Possible lower redundancy shortens the development time and therefore reduces the costs. 
\item Separation of concerns helps developers to focus on their perspective and reduces the complexity.
\item Testing and simulation can be performed on the model level. This includes automated analyses of models to prove the consistency and other properties.
\end{itemize}

An application of MDD principles to the automation domain involves the consideration of multiple disciplines with mechanical, electrical and software engineering. We had to provide models that are easy to understand and can represent the complete automation system. During the early development stages, it is helpful to visualize the conceptual idea of the automation system. The concept is then refined into a system architecture. This architecture is quite similar to a pure software system architecture in which we have to identify components and their interfaces to exchange data. Automation systems consist of actual hardware elements, e.g. robots, sensors, which have to be modeled as well as their communication channels. In a last step, we have to write the actual source code or model the system behavior based on the previously defined architecture. It is pretty straightforward that the complexity is always increasing during the development steps.

\begin{figure*}[ht]
\centering
\includegraphics[width=0.8\textwidth]{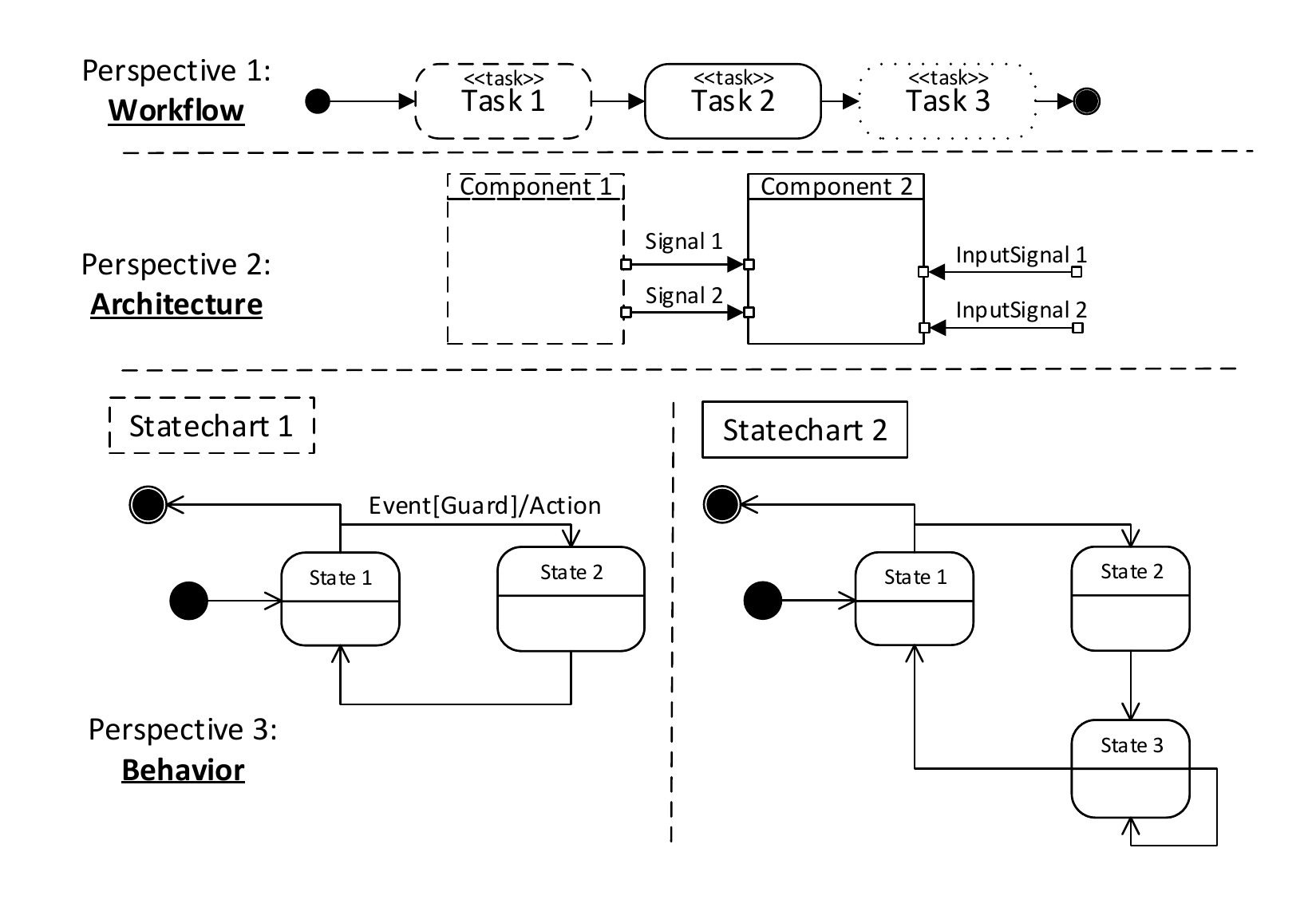}
\vspace{-10mm}
\caption{A general Example}\label{fig:3pers}
\end{figure*}

In our multi-perspective modeling approach, we followed these guidelines resulting in the three perspectives of \emph{workflow, architecture} and \emph{behavior}, now specified in more detail and pictured in Fig.~\ref{fig:3pers}:        

\paragraph{Workflow Perspective} 
The workflow perspective represents the conceptual idea of the automation system (see Fig.~\ref{fig:3pers}) capturing the technical process. Developers can describe the path of a workpiece through the automation system including the different tasks or stations that have to be performed in a specific order. Fig.~\ref{fig:3pers} shows a sequential order of three tasks. The workflow is described by UML activity diagrams and provides all standard elements such as \emph{activity, decision, merge, join, fork, initial} and \emph{final} nodes as well as transitions with possible guards, e.g. for a decision. Note that this workflow is also used to execute performance analysis, which is why all activities are equipped with arrival and service rates and transitions additionally with probability values depicting how likely it is for a workpiece to take a certain path. We refer to activity nodes as tasks throughout this paper. Tasks can be mapped to executing components of the underlying architecture. However, it is not mandatory to map all tasks to components, since some tasks may be purely mechanical parts without any implementation. In addition, one task can be mapped on multiple components simulating an execution on several machines.

\paragraph{Architecture Perspective}
The second perspective captures the logical structure of the automation system such as actual hardware parts, e.g. sensors and actuators. The communication between the machines is realized by signals that are passed between components via connectors and input/output ports enabling an information flow. External signals from the environment are also valid (see Fig.~\ref{fig:3pers}). The material flow is already represented by the workflow (see~\cite{fase}). This perspective is in style of UML composite structure diagrams. Components may have an internal behavior leading us to the third perspective.  

\paragraph{Behavior Perspective}
The behavior of architectural components is described by UML state charts with the common notation of states, transitions and their dedicated events, guards and actions. A transition with no label is by default spontaneous. Each state chart is mapped to a component on the architecture perspective. Several components can contain the same behavior, however, it is impossible that one component is described by two or more models, which would result in another variant of the system. Transitions may use signals from the architecture as events to trigger a state change, but are not limited to these events.  
 
The inter-perspective mappings are realized with a name-based mapping function at the moment. In Fig.~\ref{fig:3pers}, it is visualized by dashed and solid lines. Each perspective can contain an arbitrary amount of models. A more detailed description of the multi-perspective modeling can be found in~\cite{icse-ws,KowalAT2014}.         

\paragraph{Delta Modeling}
As a next step, we introduce delta modeling on all perspectives as well as the mapping to master the rising complexity of product lines. The general concept of delta modeling is to capture variants by means of modifications to a selected core variant~\cite{InaDelta}. Each delta contains an arbitrary number of add, remove or modify operations to transform a given model. A new variant of the system is generated by applying a set of deltas in an appropriate order to the core system. We also use deltas to manage system evolution. Integrating delta modeling allows us to perform the model modifications listed in Table~\ref*{tab:results}.

\begin{table}[t]
\centering
\small
\begin{tabular}{l|c|c|c||l|c|c|c}
\toprule
Workflow & Add & Remove & Modify &Architecture & Add & Remove & Modify \\
\midrule  
 Task	 & X	&	X			& X			& Component&X& X& X \\
  Decision/Merge	 & X	&	X	& 		& Port & X&	X &\\
   Fork/Join	 & X	&	X		& 		& Connection & X& X	&\\
 Initial/Final	 & X	&	X		& 		& Signal &&	& X \\
  Transition	 & X	&	X		& 	X	&&&	& \\
   Performance Values & 	&			& 	X	&&&	& \\
\toprule
Behavior & Add & Remove & Modify  & Mapping & Add & Remove & Modify\\
\midrule  
 State	 & X	&	X			& X	& Task $\rightarrow$ Component& X& X&	 \\
 Initial/Final	 & X	&	X		&	&  Component $\rightarrow$ State Chart & X& X&	\\
  Transition	 & X	&	X		& 	X	&&&& \\
\end{tabular}
\caption{Delta Operations}\label{tab:results}
\end{table}   
\vspace{-5mm}
\begin{figure*}[h]
\centering
\includegraphics[width=0.8\textwidth]{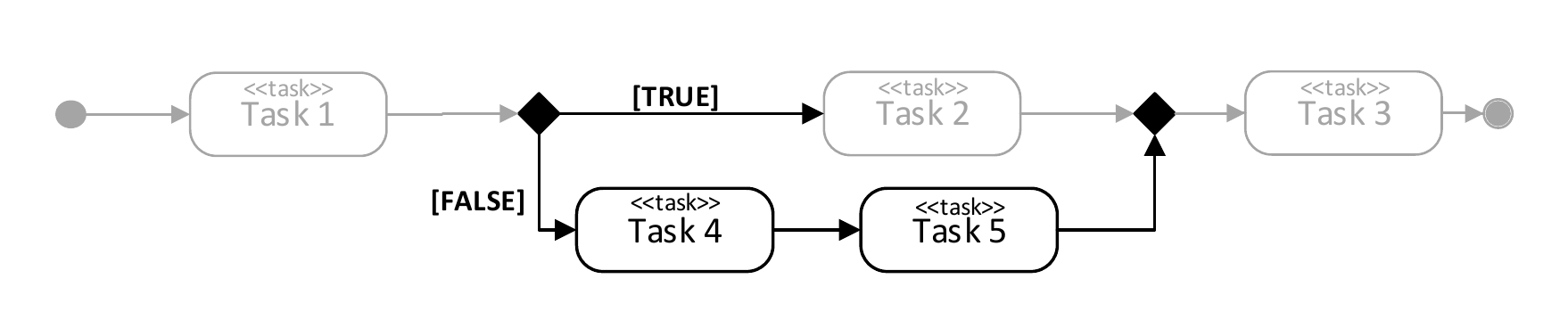}
\vspace{-5mm}
\caption{Workflow Delta}\label{fig:deltaW}
\end{figure*}
Fig.~\ref{fig:deltaW} shows a basic delta for the first perspective introducing an alternative path in the system. Bold lines indicate the performed changes. The process is similar for each perspective and the mapping. 

During development, we must ensure consistency on each perspective in isolation and their mappings respectively. If inconsistencies occur in the process, the developer should be informed about the problems and supported in their solution. While this is quite similar to common IDE features, e.g., automatically adding import-statements for used classes in Eclipse, we also have to ensure that all automatically generated variants are valid, which gets increasingly more difficult in large product lines. The consistency checking was a completely manual task so far. Developers had to search the generated models for inconsistencies, which was a tedious process given at least three models per variant~\cite{icse-ws,KowalAT2014}. This work lifts this limitation by integrating a constraint-based consistency checking method, which is easily extendable, automatically checks all generated variants, supports the development process and works in an incremental fashion to reduce performance costs.      

\section{Consistency Checking in Delta-oriented UML-Models}
The Unified Modeling Language (UML) is the most frequently used specification and the de facto standard modeling language for software system development~\cite{UMLbible}. Overall, there are 14 model types that can be used to describe different perspectives or abstraction levels of a system, e.g., behavior or design, reducing the level of complexity and supporting the distribution of responsibilities among stakeholders. Using multiple UML models during the development process ultimately results in a strong dependency between them making consistency an important aspect to ensure a valid system in the end. However, UML does not provide us with a formal notation. Inconsistencies may already occur during the development of one model. This aspect is intensified by consideration of multiple models for different perspectives~\cite{Ibrahim2011}. Inconsistencies can easily lead to errors in the software system~\cite{Huzar2005,1620102}. In Fig.~\ref{fig:3pers} and more specific state chart~1, a removal of the transition from state~2 to state~1 results in a deadlock that would require a shutdown of the operating automation system. Hence, it is paramount to identify inconsistency as early as possible, e.g., during the design phase, and to fix them~\cite{Spanoudakis01inconsistencymanagement}. The detection of inconsistencies requires a set of consistency rules, which we discuss in the following.    

\subsection{Consistency Rules}
Some specifications for consistency rules can be found in the UML standard and are referred to as rules for well-formed models~\cite{omg}. Most consistency rules defined in the literature are related to UML class diagrams, which are not present in our multi-perspective modeling approach. In general, the consistency rules found in the UML standard or other literature help us to ensure consistency on each perspective in isolation. For our modeling approach, it is also mandatory to ensure consistency across the three perspectives as well as for the integrated delta modeling approach in an efficient manner. We identified three mandatory rule categories for our approach with:
\begin{enumerate}
\item Intra-perspective rules affect each perspective in isolation. (\textit{Level 1})
\item Inter-perspective rules affect one full variant of the system. (\textit{Level 2})
\item Cross-variant rules affect the full product line. (\textit{Level 3})
\end{enumerate}  
It is obvious that rules in the first category must be enforced to develop a valid variant of the system as well. Also, intra- and inter-perspective rules must be fulfilled in addition to the cross-variant rules to ensure a completely valid product line. 

The workflow perspective is represented by UML activity diagrams. We identified and implemented 22 rules that have to be fulfilled in order to receive a valid workflow. The complexity of the rules varies from proving the existence of an initial or final node to reachability checks for each node. An architecture as well as behavior model must comply to 11 different consistency rules such as each port must have a connector or a state chart has exactly one initial state. The difference in numbers can be explained by a significantly larger number of model elements at the workflow level. Hence, we must validate a set of 44 different consistency rules for \textit{Level 1}. The mapping between the perspectives can be validated by using an additional number of 5 rules, e.g. each state chart is connected to at least one component in the architecture. Finally in \textit{Level 3}, each delta is checked for 3 rules to ensure general applicability. A model must exist in order to apply a delta meaning that we have to check for its existence first. As a result, we have one rule for each perspective. However, each delta requires to recheck the previous levels subsequently, since it may alter the perspectives (\textit{Level 1}) or the mapping (\textit{Level 2}). Each perspective may contain an arbitrary number of models. Given this set of rules, it is possible to use different validation concepts, whereas each concept has another impact at the actual number of checks performed to enforce validity across the whole product line. We discuss three techniques in the following and show the evaluation results of each technique in the next section.            
           
\subsection{Product-based Consistency Checking}
A product-based technique completely analyzes all variants of the product line. All checks are executed for each individual variant without considering the variability information contained in the deltas. 

In case of the core variant, we can directly support the developer during its development by continuously checking all perspectives and the mapping in our implementation. However, this is not feasible for additional variants taking delta modeling into account in which several deltas may depend on each other and can be combined in multiple ways. The user specifies the desired variants based on the selection of deltas and their application order. Based on this information, we can automatically validate all variants meaning the core and all selected deltas given our set of consistency rules. The worst case scenario would be a product-based method at this point~\cite{acta}. First of all, this solution would check if the core variant is free of inconsistencies neglecting the fact that our actual implementation already ensures a consistent core for now. This process is repeated for each generated variant resulting in lots of redundant checks due to the fact that most deltas do not change each model and the mapping. E.g. the workflow may never be touched, but is again validated for each variant. Such a brute-force approach gets increasingly less efficient with a growing number of variants.       

\subsection{Product-based Incremental Consistency Checking}
A product-based incremental approach also works on all generated variants of the product line. However, it reuses the knowledge of the modifications stored in a delta to improve the consistency checking. Since the deltas provide differences between two variants, we only have to recheck models that have been altered giving us smaller increments.    

Although the second approach still bears a product-based concept as foundation, it is superior due to its incremental nature. The principle of delta modeling itself provides us with the necessary optimization potential. Each delta encapsulates the modifications executed on the specified model in the respective perspective. Taking this information into account, we can improve the product-based method in two aspects: First, we do not have to recheck models that are already valid. After the initial validation of the core variant, we can safely assume that all perspectives and their mappings are consistent. Most likely a delta does not touch all perspectives, e.g., it only modifies one perspective, making consistency checks on the remaining perspectives obsolete. Second, some changes do not even require another iteration on the modified model, e.g., modifying the performance values in the workflow. The decision to recheck a specific model always depends on the types of modifications in the delta. Based on the transformations that are possible in deltas (see Table~\ref{tab:results}) and the defined consistency rules, we are able to decide if any rule may be violated by the transformation and, therefore, if a revalidation of the model is necessary. As a result, we can safely expect a significant reduction of the consistency checks compared to the pure product-based approach, since we avoid many redundant checks. However, this is only the case in small deltas that are limited to some models. A large delta affecting all models and the mapping would result in a validation of the complete variant, and we would lose the incremental benefits here.             

\subsection{Fully Delta-based Incremental Consistency Checking}
The next step is an incremental approach taking full advantage of the information stored in a delta. It is possible to validate only certain parts of a concrete model based on the change in a delta. For example, if we consider the addition of a task and the necessary reachability check in our workflow. This task must be reachable from at least one initial node and we must reach at least one final node to ensure consistency. The product-based incremental method would recheck the reachability of all tasks, since it checks complete models, which again contains redundancy. It is possible to do this check only for the added task. However, more complex cases are not uncommon, especially considering the removal of elements. Simply removing a transition can result in many inconsistencies with regard to reachability. It is paramount to scan the full delta first and narrow the affected model elements down, before a decision for specific consistency checks is reasonable. We found several of such cases based on the change operations and their application contexts. As a result, we minimize the chance to enter the worst case scenario of a full revalidation which is highly unlikely using this method. In general, we can expect an even further reduction of redundant checks compared to the product-based incremental approach.   

Both product-based approaches are fully functional in our implementation, while the fully delta-based method is at an experimental stage, since not all patterns have been identified at the moment. 

\section{Case Example}
The following section is divided into two parts. First, we explain the automation system in more detail on which we applied the multi-perspective modeling approach. Second, we present the results of the three consistency checking methods that were separately used to ensure the generation of valid variants. As a result, we should observe a reduction of necessary consistency checks from the product-based to the fully incremental approach.  

\subsection{The Pick and Place Unit - An Automation System}
\begin{figure}[ht]\centering
\includegraphics[width=0.4\columnwidth]{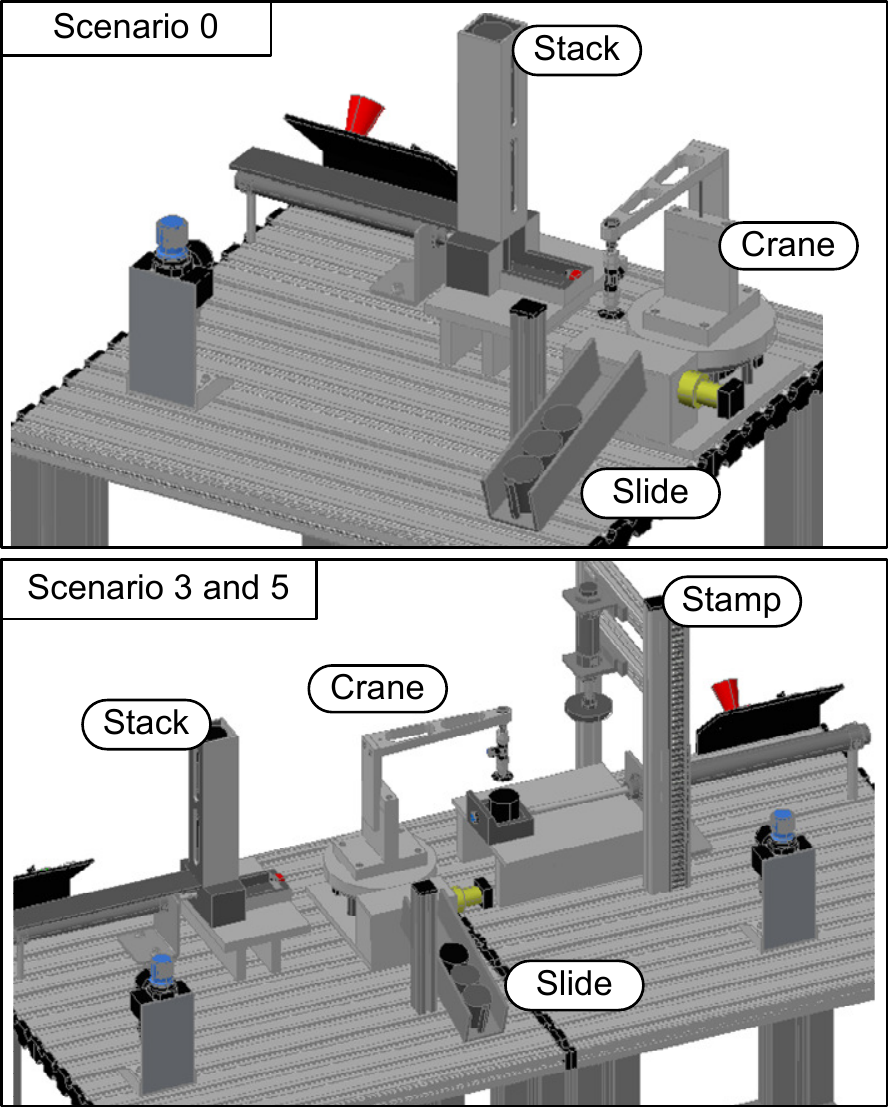}
\caption{Mechanical setup of the PPU~\cite{icse-ws}}\label{fig:ppu}
\end{figure}

The Pick and Place Unit (PPU) is a universal production demonstrator to study evolution in the automation domain~\cite{CL-IECON2013}. In its current setup, it consists of 15 different variants (referred to as scenarios) simulating a product line. Documentation, simulation files and source code are available to the public~\cite{ppu}. As a case example, three out of the fifteen variants are used (see Fig.~\ref{fig:ppu}) and briefly described in the following.

\begin{figure}[hp]\centering
\includegraphics[width=1.0\columnwidth]{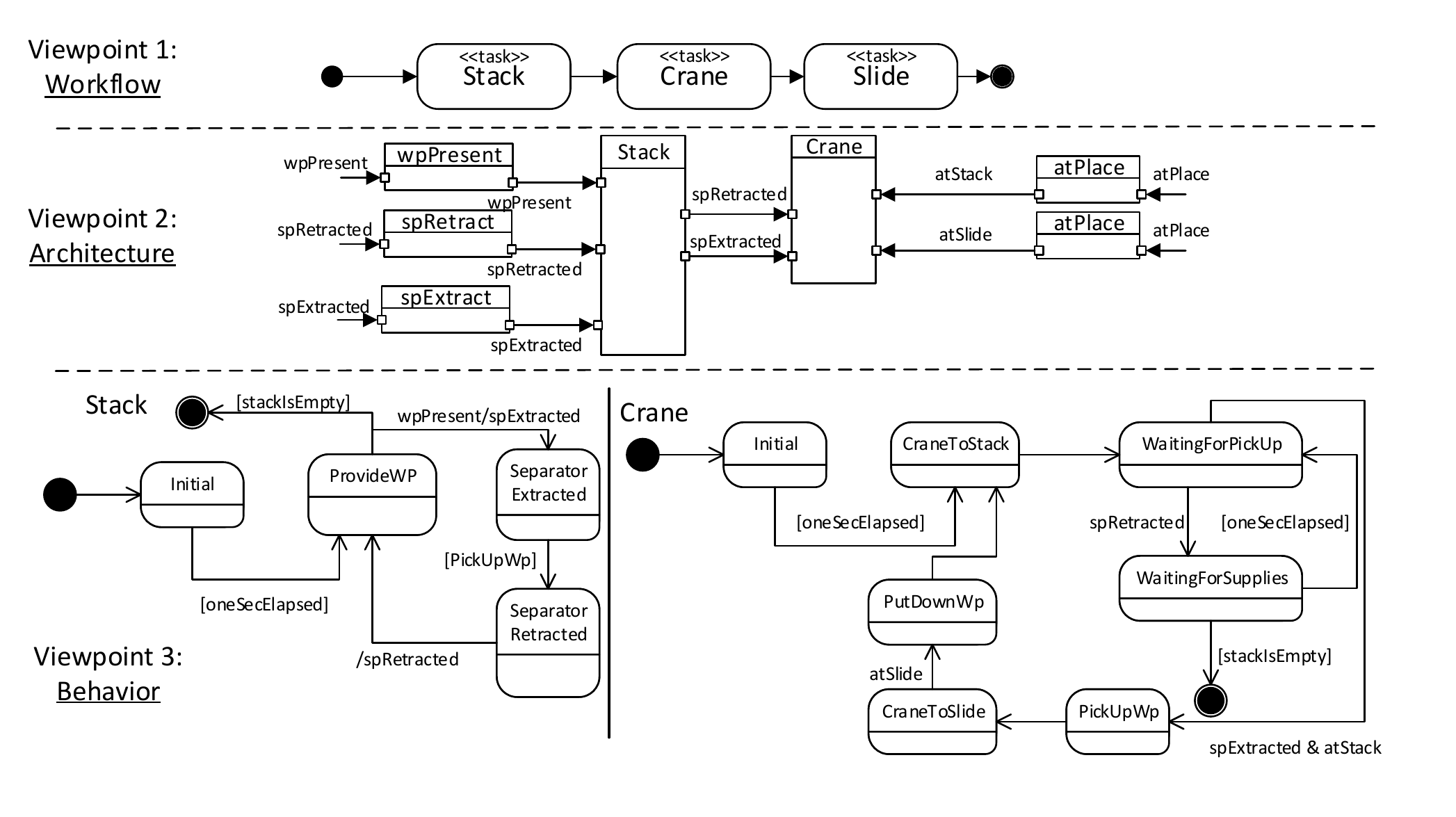}\vspace{-5mm}
\caption{Scenario 0~\cite{icse-ws}}\label{fig:sc0}
\includegraphics[width=1.0\columnwidth]{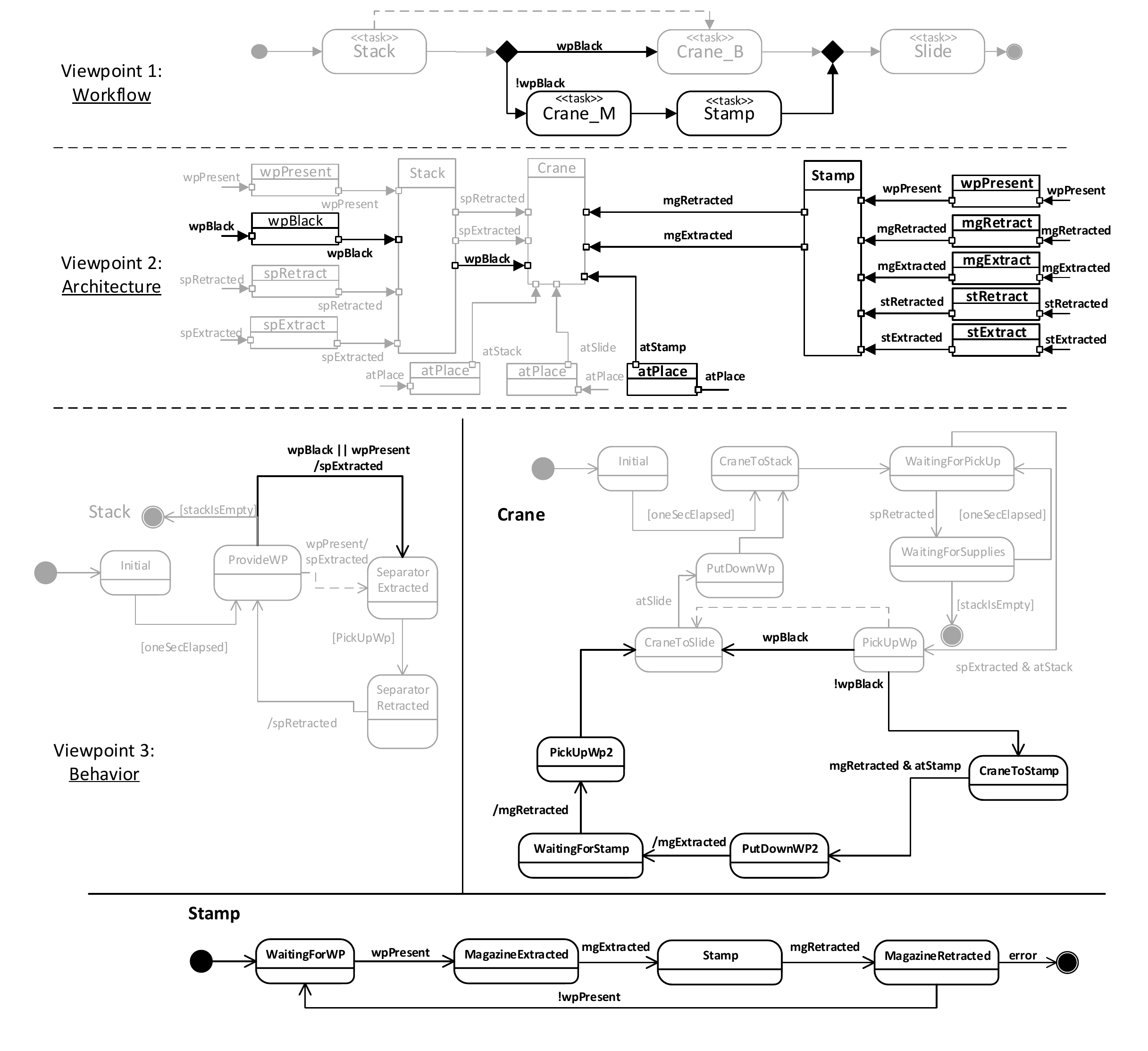}\vspace{-5mm}
\caption{Scenario 3~\cite{KowalAT2014}}\label{fig:sc3}
\end{figure}

The PPU consists of three basic components in scenario 0: a storage component providing workpieces called \textit{stack}, a mechanical \textit{slide} simulating the output entity of the system, and a \textit{crane} for the transportation of the cylindrical plastic workpieces between the other two components. Fig.~\ref{fig:sc0} shows the scenario modeled within our multi-perspective modeling approach. The workflow has three tasks abstracting the transportation route of workpieces as already described. Two tasks, \textit{stack} and \textit{crane}, are mapped to a component on the second perspective. The \textit{slide} does not need one, since it is only a piece of metal without any additional functionality. Each component gets further information from different sensors, e.g. \textit{wpPresent}. The internal implementation is described by the two state charts that are again mapped to the two components. The mapping is visualized with the help of identical names.        

In scenario~3, a second workpiece type is introduced that has to be processed in a different manner. The new metallic workpieces have to stamped before they are transported to the slide. As a result, we create an alternative path in our workflow. It is important to mention that both crane tasks are mapped onto the same component in the architecture (see Fig.~\ref{fig:sc3}) meaning that we still have one physical \textit{crane}. The new stamping module is also represented in the architecture as well as several new sensors providing additional information about positioning and the workpiece type. Hence, we must also introduce a new state chart for the \textit{stamp} and extend the old crane behavior to handle metallic pieces. 	  

Scenario~5 is purely a software optimization. In scenario~3, the \textit{crane} waits for the \textit{stamp} to finish the process and continuous the transportation afterwards. This idle time is now used to transport another plastic workpiece to the slide. Afterwards, the \textit{crane} returns to the \textit{stamp} and continuous processing the metallic workpiece.

The first scenario serves as our core variant in the evaluation. Overall, we have exactly three variants, since the application of the delta for scenario~5 is only reasonable if we generated the third scenario beforehand making it a very small delta. Note that Fig.~\ref{fig:sc0} and Fig.~\ref{fig:sc3} are only extracts from the full variant, since only the state charts for the main components, \textit{stack}, \textit{crane} and \textit{stamp}, are depicted in this paper.    

\subsection{Evaluation}
We applied our approach to the three scenarios of the PPU that were described before. All necessary models were already at hand due to previous work~\cite{icse-ws,KowalAT2014}. Goal of the present evaluation was to find any inconsistencies in these models and to compare the different consistency checking methods. A check is defined as the validation of exactly one rule in a specific model. Consider, e.g., the reachability in state charts, which has to be fulfilled for all states giving us one check. 

First of all, it is mandatory to validate the core variant which is scenario~0. Overall, the core contains 9 different models that have to be checked resulting in 85 individual consistency checks. We tested the core variant on the physical machine and were not able to deduce any errors or general problems. Nevertheless, our consistency checking found 19 inconsistencies in our core. It is possible to have multiple inconsistencies for one rule, e.g., depending on how many states are not reachable. An analysis of the errors exposed that the majority is related to the architecture perspective and concerns signals that are not used in any state chart or components that are not mapped to any task, e.g., the localization sensors. These problems are not crucial to guarantee a safely working PPU and do not affect the overall behavior of the variant, but we improved our models accordingly. These 85 checks remain the same for each consistency checking method, since it is always mandatory to validate the core completely. 

The delta to reach scenario~3 of the PPU contains an additional state chart and performs changes on each perspective (see Fig.~\ref{fig:sc3}). The product-based approach generates the final variant first and rechecks all models, although several state charts for the sensors are never modified. As a result, we have to perform all 85 checks again as well as 12 additional ones. Surprisingly, we got a large increase in the number of inconsistencies with a total of 65. They range from unused signals in the architecture to duplicate state names in the behavior. Since, even the rather small example of the PPU can result in so many inconsistencies, we find it imperative to provide consistency checking methods in our multi-perspective modeling tool. The product-based incremental approach reduces the number of checks to 61 for scenario~3, which is a reduction of about 30\%. If we consider a small delta such as for scenario~5, the product-based incremental approach gets increasingly more beneficial. Again, we need 97 checks for the product-based one, since we do not add a completely new model. However, the incremental approach only needs 38 checks. This is due to the fact that we modify the workflow and only one state chart model, the \textit{crane}, while all other models, e.g., architecture, remain identical. All approaches always found the same inconsistencies. 

The delta-based incremental approach should reduce the number of checks even further. However, we only have preliminary results at this point which is why we exemplarily explain the results for the \textit{crane} behavior in Scenario~5. Both product-based approaches would need 11 checks of the full model to ensure consistency. By using the full information stored in a delta, we can reduce the number to 6. For example, we can scan a delta if a second initial node is added which is forbidden in state charts neglecting parallel state machines here.

\section{Implementation}
We currently have two separate prototypical implementations of the multi-perspective modeling approach. The first uses textual domain-specific languages (DSLs) realized with Xtext\footnote{\url{http://www.eclipse.org/Xtext/}}. XText is continuously improved by a large community and an open-source framework. Each DSL in our tool underlies a grammar that already prevents some basic inconsistencies. However, as previously described it was necessary to introduce a large number of consistency rules separately. In addition, XText provides us with an easy solution to implement IDE comfort functions such as syntax highlighting or auto-completion. The tool can automatically generate a specific variant of an automation system by applying the necessary deltas to the designated core, e.g. scenario~0. The consistency methods were tested with this implementation. Therefore, we can support the user in finding and solving inconsistencies in the models. In addition, we can export a specific variant to CODESYS\footnote{\url{http://www.codesys.com/}}. CODESYS supports the IEC~61131 standard and is widely used in industry to develop automation control software. The CODESYS interface is based on XML of which we take advantage in our export. We modeled the core variant and the deltas for scenario~0 and 3 respectively for our evaluation. After proving that no inconsistencies remained in our models, we exported the variants to CODESYS and tested our models on the real PPU. Without the consistency checking step it would be possible to damage physical components or end up in a deadlock. To our relief all tests were successful and nothing was damaged.   

The second implementation relies on graphical models. It is realized with the Graphical Modeling Framework (GMF)\footnote{\url{http://www.eclipse.org/gmf-tooling/}} and the Graphical Editing Framework (GEF)\footnote{\url{https://eclipse.org/gef/}}. We can model each perspective and the mapping with the help of graphical editors. However, the consistency checks only work at the textual level, although a model-to-model transformation between both implementations is possible due to the nature of the Eclipse Modeling Framework (EMF)\footnote{\url{https://eclipse.org/modeling/emf/}} meta-models underlying both implementations.

\section{Related Work}
In the following, we do not review prior work regarding variability and delta modeling, which is extensively done in~\cite{icse-ws,fase}. Instead, we solely focus on the presented extension of consistency checking in UML-models. 
UML is often the de-facto standard in industry to develop software systems and simultaneously model-to-code generators get more and more popular as they reduce development efforts~\cite{4721312}. The correctness of generated code heavily depends on consistency in the UML-models which is why consistency techniques are a key aspect in MDD~\cite{4721312,Thum:2014:CSA:2620784.2580950}. Usmann et al. identified in their survey five consistency types in UML-models with inter-model, intra-model, evolution, semantic and syntactic consistency. The types are based on a literature review concerning existing UML consistency checking methods. Most of the methods are based on class-, sequence- and state chart models leaving our architecture and activity perspectives excluded. Nonetheless, we adapted the general consistency types in our work. A more recent study supports the results of the survey~\cite{Torre:2014:UCR:2601248.2601292}. Additional classification of inconsistencies can be found in literature~\cite{Mens05aframework}. 

Egyed et al. have done an extensive amount of research to ensure consistency in UML-models. In~\cite{Egyed:2006:ICC:1134285.1134339}, an instant consistency checking method working on the most three most popular UML-models (see previous paragraph) has been proposed and implemented in a tool called UML/Analyzer which is integrated within the famous IBM Rational Rose$^{TM}$. An incremental checking is not necessary instant and often requires additional declarations~\cite{Egyed:2006:ICC:1134285.1134339}. We realized a combination of both approaches is most beneficial in our case. Our core is instantly validated during development. However, this is not possible for the variability, which is not considered in~\cite{Egyed:2006:ICC:1134285.1134339}. Our consistency checking works incrementally in terms of variability while no additional declarations in the models are necessary.

Software product lines are often represented with the help of feature models depicting commonalities and variability of the system. Egyed et al. reuse this knowledge to provide consistency for lower level UML-models~\cite{Lopez-Herrejon:2010:DIM:2164046.2164064}. A similar approach is presented in~\cite{Demuth:2011:CMT:2025113.2025189}. This aspect is still missing in our approach, since we do not have any high-level feature model to which our multi-perspective models as well as the deltas can be bound. This improvement is left for future work.               

\section{Conclusion}
We have presented multiple consistency checking methods based on our multi-perspective modeling approach and evaluated them on a real-world automation system. All models were created using our textual DSLs implemented with XText. The variability modeling approach at hand was delta modeling providing us with the foundation to optimize a product-based consistency approach. We defined three different categories with intra-, inter-perspective and cross-variant rules for our UML-models and integrated all of them in our tool chain. Based on these rules and three variants of the PPU, we evaluated different types of consistency checking approaches with an ever increasing performance in terms of number of executed checks. It was surprising to find this many inconsistencies in the variants, although they successfully ran on the physical machine and we only considered rather small models due to the size of the PPU. This aspect endorses us that consistency checks in models are necessary. However, the delta-based incremental approach still needs further investigation, but the first results are promising. With the extension of inconsistency detection, we now can fully support the development process of an automation system from the design phase to the actual execution on the machine, since our tool supports code generation of a specific variant for the automation control software CODESYS. 
 
Further investigations are necessary towards an applicability in the industrial domain, especially concerning larger case studies. The next level after finding the inconsistencies would be to provide repair operations that can be selected by the developer, similar to standard IDE features as available in Eclipse. In addition, the graphical editors are standalone and require the same consistency checking concept as the textual version. An extension to feature models also is a possible future step. We can reuse the variability information in feature models to simplify the variant generation process for the developers by selecting features instead of specific deltas.

\bibliographystyle{eptcs}
\bibliography{work_report}
\end{document}